\documentclass{PoS}

\title{Finite Size Scaling for the O(N) universality class from Renormalization Group Methods}

\ShortTitle{Finite Size Scaling for the O(N) universality class from Renormalization Group Methods}

\author{\speaker{Bertram Klein}\\
        Institut f\"ur Theoretische Physik T39, Physik Department, Technische Universit\"at M\"unchen\\
        E-mail: \email{bklein@ph.tum.de}}

\author{Jens Braun\\
        Theory Group, TRIUMF\\
        E-mail: \email{braun@triumf.ca}}

\abstract{The QCD phase diagram at finite temperature and density is a topic of considerable interest. 
Although much progress has been made in recent years, some open questions remain.
Even at zero density, the order of the transition for two light flavors of fermions has not yet been conclusively established. While considerable evidence exists in favor of a second-order transition for massless quarks and a crossover for massive quarks, some recent results with two flavors of staggered fermions suggest a transition of first order.

Since lattice simulations are performed in finite simulation volumes, actual phase transitions cannot be observed directly. Thus, finite-size scaling is a very useful tool in the analysis of lattice data. By comparing the scaling behavior of observables to the expected scaling properties, values of critical exponents can be confirmed and the order as well as the universality class of a transition can be established. In the comparison to lattice QCD results, the critical exponents and the universal scaling functions have been obtained mainly by means of lattice simulations of O(N) spin models, and results are usually restricted to the critical temperature or the point at which the susceptibilities peak.  

We propose to use a non-perturbative Renormalization Group method for this purpose.
We have calculated the critical finite-size scaling behavior and the universal scaling functions for the three-dimensional O(4)-model for a wide range of temperatures and values of the symmetry breaking parameter. Our results are suitable for a comparison to lattice QCD results for the chiral susceptibility and the order parameter and can be used to check the consistency of the finite-size scaling behavior with that of the O(N) universality class.}

\FullConference{The XXV International Symposium on Lattice Field Theory\\
		 July 30 - August 4 2007\\
		 Regensburg, Germany}

\begin{document}

\section{Introduction}

In QCD, two phase transitions take place at finite temperature and density, a deconfinement phase transition dominated by the gauge fields, and a chiral symmetry restoration transition. 
Whether these transitions coincide is still not settled \cite{Aoki:2007xx, Aoki:2007xy, Endrodi:2007xx}.
If the QCD phase transition is dominated by the restoration of chiral symmetry, the conventional scenario 
\cite{Pisarski:1983ms} holds that the transition is of second order for $N_f=2$ massless quark flavors, and  a crossover for massive quarks. While there is much evidence in favor of this scenario \cite{Aoki:2006we}, results obtained with two dynamical flavors of staggered fermions suggest a first-order transition \cite{D'Elia:2005bv}, and recent work by the same group supports these findings \cite{Cossu:2007mn, Cossu:2007xx}. 

Since lattice simulations are necessarily performed in finite simulation volumes, and since strictly speaking phase transitions only occur in the infinite-volume limit, it is difficult to conclusively determine the order of a phase transition. A further complication is the explicit breaking of the chiral symmetry by the finite quark mass, which corresponds to an external symmetry-breaking field.
An important tool is therefore the finite-size scaling analysis of the simulation data. Due to universality, in the case of critical behavior results are expected to fall onto universal scaling curves, characterized by critical exponents and scaling functions of a particular universality class.

Necessary for establishing the nature of critical behavior in a finite volume is therefore knowledge of two things: of the critical exponents, and of the scaling behavior.
So far, scaling functions have mainly been determined from lattice simulations of O(N) spin models \cite{Engels:2001bq, Schulze:2001zg}. 
While there is evidence of scaling for Wilson fermions, the expected infinite-volume scaling behavior is not seen for staggered fermions \cite{Mendes:2006zf, Mendes:2007xx}.
Results with a modified QCD action with two flavors of staggered fermions suggest that current simulation volumes might be outside the scaling region \cite{Kogut:2006gt}.

In the work presented, we use a non-perturbative Renormalization Group (RG) method to calculate critical exponents as well as the finite-size scaling functions. While an abundance of results exists on critical exponents, this is to our knowledge the first time that finite-size scaling functions have been obtained with non-perturbative RG methods. As an advantage, a very wide range of parameters and volume sizes is accessible to the method, and thus direct comparisons to lattice simulations are possible and the size of the scaling region can be assessed.

\section{Renormalization Group method}
We use a non-perturbative Renormalization Group method to calculate observables such as the order parameter and the susceptibility. The method accounts for long-range fluctuations and is suitable to describe critical behavior. For a review of functional RG methods, see e.g. \cite{Berges:2000ew, Pawlowski:2005xe, Gies:2006wv}.

In the framework of the effective action, an infrared cutoff $k$ is introduced,  and by varying the cutoff scale quantum fluctuations are integrated out in a systematic way. The values of coupling constants at an initial UV cutoff scale $\Lambda$ are the input. After the IR cutoff has been lowered to $k=0$, all quantum fluctuations below $\Lambda$ have been integrated out and are included in the values of the couplings in the effective action. The change of the couplings under a variation of the cutoff scale $k$ is governed by a flow equation. For a particular choice of the RG cutoff scheme, the flow equation for the $d$-dimensional O(N)-model in infinite volume reads \cite{Litim:2001up, Litim:2001hk, Bohr:2000gp}
\begin{equation}
k \frac{\partial}{\partial k} U_k(\sigma, \vec{\pi}) = \frac{(k^2)^{d/2+1}}{4 \pi \;\Gamma(d/2+1)}\left(\frac{(N-1)}{k^2 + M_\pi^2(k)} + \frac{1}{k^2+ M_\sigma^2(k)}\right).
\end{equation}   
We have adapted this cutoff scheme to calculations in a finite volume \cite{Braun:2005fj},
and are currently investigating the dependence of our finite-volume results on the choice of cutoff scheme.
The effective potential is expanded in local $n$-point couplings around its minimum $\sigma_0(k)$
\[
U_k(\sigma, \vec{\pi}) = a_0(k) + a_1(k) (\sigma^2 + \vec{\pi}^2 - \sigma_0(k)^2) + a_2(k) (\sigma^2 + \vec{\pi}^2 - \sigma_0(k)^2)^2 + \ldots - H \sigma,
\]
where $H$ is the fixed, external symmetry-breaking field, and the additional condition $ 2 a_1(k) \sigma_0(k) = H$ keeps the minimum at $(\sigma, \vec{\pi}) = (\sigma_0(k), \vec{0})$. The RG flow equation is solved numerically.

In $d=3$ dimensions, the initial value of the minimum at the cutoff scale $\sigma_0(\Lambda)$ serves as proxy for the temperature, $(\sigma_0(\Lambda) - \sigma_0^{\mathrm{critical}}(\Lambda)) \sim (T - T_c)$.
For this first calculation, we have chosen a cutoff scale $\Lambda = 1.0$ GeV, which is comparable to a typical lattice cutoff ($\pi/a \approx 1.5$ GeV) in current lattice simulations.

\section{Scaling in Infinite Volume}

Close to a critical point, such as a second-order phase transition, the correlation length $\xi$ diverges, and long-range fluctuations dominate the behavior of a system. As a consequence, certain observables become independent of the short-scale details of the system. Close to the transition, their behavior is described by only a few critical exponents and scaling functions which  are characteristic of the universality class of the transition. For example, the order parameter behaves according to
\begin{equation}
M(t, h) = h^{1/\delta} f(z), \;\;\; z = t/h^{1/\delta},
\end{equation} 
where $f(z)$ is the scaling function and $z$ is the scaling variable.
The two conditions $M(t, h=0) = (-t)^\beta$ and $f(0)=1$ determine normalization constants for the dimensionless temperature and field parameters,  $t= (T-T_c)/T_0$ and $h=H/H_0$.

\begin{table} 
\begin{tabular}{llllll} 
\hline
&& $\phantom{(}\beta$ & $\nu$ & $\phantom{(}\eta$ & $\phantom{(}\delta$ \\
\hline
J. Engels {\it et al.} \cite{Engels:2001bq} & lattice & $\phantom{(}0.380$ &  $0.7423$ & $(0.024)$ & $\phantom{(}4.86$ \\
D. F. Lititm and J. M. Pawlowski \cite{Litim:2001hk} & RG & $(0.4022)$ & $0.8043$ & $-$ & $(5.00)$\\
this work 	& RG & $\phantom{(}0.4030(3)$ & $0.8053(6)$ & $(0.0046(4))$ & $\phantom{(}4.9727(5)$\\
\hline
\end{tabular} 
\caption{Critical exponents for O(4) in $d=3$. 
Values in brackets are calculated 
using scaling laws.} 
\label{tab:critex} 
\end{table} 
We obtain values for the critical exponents $\beta$, $\nu$, and $\delta$ by fitting directly to the observables $M =  \sigma_0$ and $\xi = 1/M_\sigma$ as functions of the temperature and the symmetry-breaking field. In Tab.~\ref{tab:critex}, our results are compared to those of a different RG calculation, 
where we find perfect agreement, and lattice spin model calculations, where the agreement is reasonably good and the deviations can likely be explained by the restriction to local couplings in our approach. 

\begin{figure}
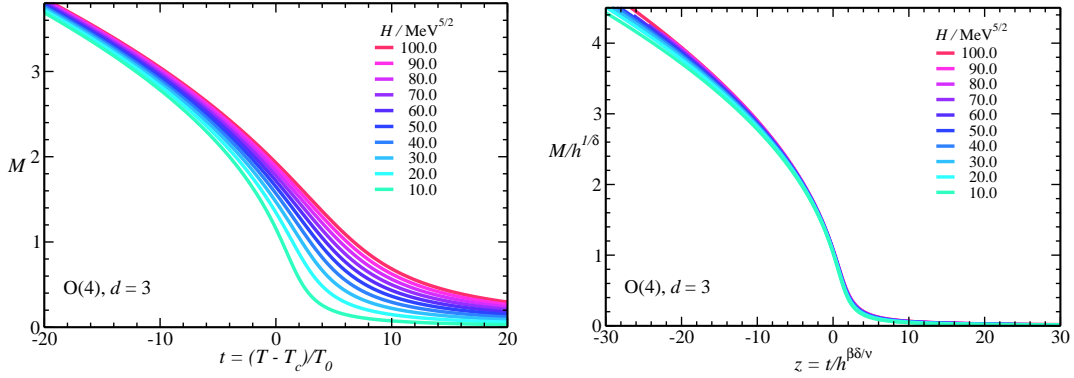
 
\vspace*{5mm}
\begin{center}
\includegraphics[width=.45\textwidth]{d3n4a3-2orderparameter_normalized-vs-t-crange10_0-100_0-line.eps} \hspace{2mm}
\includegraphics[width=.46\textwidth]{d3n4a3-2orderparameter_scaled-vs-t-crange10_0-100_0-line.eps} 
\end{center}
\caption{Scaling for the order parameter $M$ in infinite volume. $M$ is shown as a function of the temperature $t$ for several different values of the symmetry breaking field $H$ (left panel). After rescaling $M/h^{1/\delta}$ as a function of $z$, results for these values of $H$ collapse onto a single curve (right panel).}  
\label{fig:IVscaling} 
\end{figure}

\begin{figure}
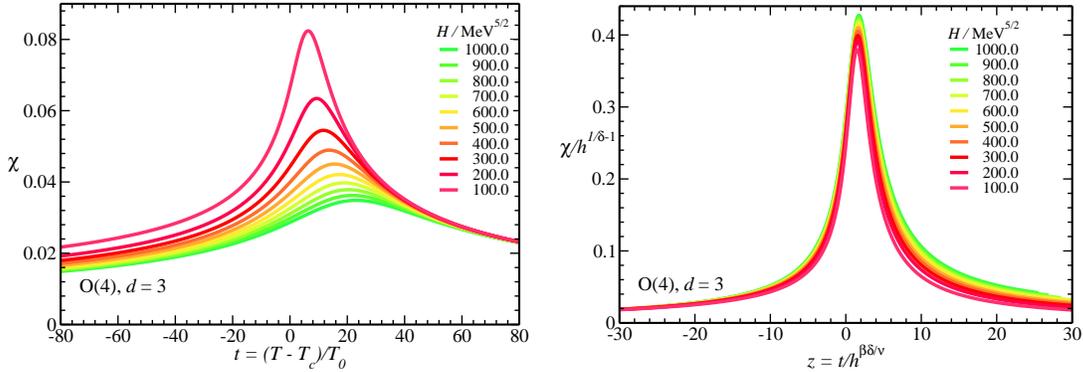
 
\vspace*{7mm}
\begin{center}
\includegraphics[width=.46\textwidth]{d3n4a3-2susceptibility_normalized-vs-t-crange100_0-1000_0-line.eps} \hspace{2mm}
\includegraphics[width=.46\textwidth]{d3n4a3-2susceptibility_scaled-vs-t-crange100_0-1000_0-line.eps} 
\end{center}
\caption{Scaling for the susceptibility $\chi$ in infinite volume. $\chi$ is shown as a function of the temperature $t$ for several different values of the field $H$ (left panel). After rescaling $\chi/h^{(1/\delta-1)}$ as a function of $z$, results for these values of $H$ collapse onto a single curve (right panel).}  
\label{fig:IVsuscscaling} 
\end{figure}

In Figs.~\ref{fig:IVscaling} and \ref{fig:IVsuscscaling}, we validate the results for the critical exponents by 
confirming the scaling behavior. In Fig.~\ref{fig:IVscaling},the order parameter $M$ is shown as a function of the temperature $t$ for several different values of the field $H$ (right).  Plotting $M/h^{1/\delta}$ against the scaling variable $z$, the different curves collapse onto the universal scaling function $f(z)$ (left). 
In the second figure, scaling is validated for the susceptibility, again the curves collapse after rescaling.
We have checked that the scaling behavior holds over three orders of magnitude for the field, from $H =1.0$ to $1.0 \times 10^3$ MeV$^{5/2}$. 
This confirms that our approach incorporates the critical fluctuations responsible for the universal behavior.

\section{Finite-Size Scaling}

\begin{figure}
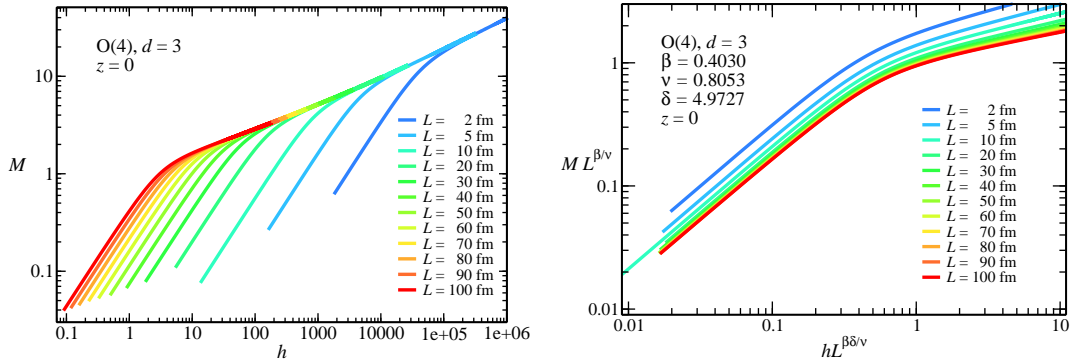
 
\vspace*{5mm}
\begin{center}
\includegraphics[width=.46\textwidth]{d3n4a3-2orderparameter_normalized_lattice2007-line.eps} 
\hspace{2mm}
\includegraphics[width=.448\textwidth]{d3n4a3-2orderparameter_FVscaled_lattice2007-line.eps} 
\end{center}
\caption{Finite-size scaling for the order parameter $M$ at $z=0$. $M$ is shown as a function of $h$ for different volume sizes (left). The finite-size scaled order parameter $M L^{\beta/\nu}$ is plotted against the finite-size scaling variable $hL^{\beta \delta /\nu}$ (right).}  
\label{fig:orderFVscaling} 
\end{figure}

\begin{figure}
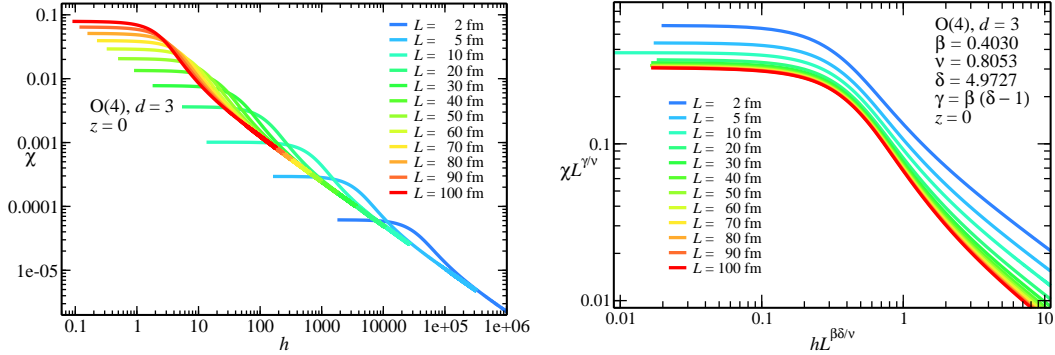
 
\vspace*{5mm}
\begin{center}
\includegraphics[width=.46\textwidth]{d3n4a3-2susceptibility_normalized_lattice2007-line.eps} 
\hspace{2mm}
\includegraphics[width=.435\textwidth]{d3n4a3-2susceptibility_FVscaled_lattice2007-line.eps} 
\end{center}
\caption{Finite-size scaling for the susceptibility $\chi$ at $z=0$.   $\chi$ is plotted vs. $h$ for different volume sizes (left). The finite-size scaled susceptibility $L^{\gamma/\nu} \chi$ is plotted vs. the finite-size scaling variable $hL^{\beta \delta /\nu}$ (right).}  
\label{fig:suscFVscaling} 
\end{figure}

Since a finite volume provides a natural IR cutoff, the long-range fluctuations are affected and the critical behavior is changed when a system is put into a finite volume. Because these cutoff effects depend on the relative size of the correlation length $\xi$ and the volume size $L$, it is natural to assume that deviations from the infinite-volume scaling behavior are controlled by this ratio. 
According to the finite-size scaling hypothesis \cite{Fisher:1971ks}, the ratio of a thermodynamic observable in the finite-size and the infinite-size system at the same temperature is a function of \emph{only} this ratio $\xi/L$.  
For example, according to this hypothesis one expects for the order parameter in finite and infinite volume
\begin{equation}
\frac{M_L(t)}{M_\infty(t)} = {\mathcal F}\left(\frac{\xi(t)}{L}\right).
\end{equation}
If this hypothesis holds, universal finite-size scaling functions can be obtained by realizing that one needs to change $t \sim L^{-1/\nu}$ with the volume to keep the ratio $\xi/L$ constant, since $\xi \sim t^{-\nu}$. 
To keep the scaling variable $z=t/h^{1/(\beta \delta)}$ constant as well, one also needs to vary $h \sim L^{-\beta\delta/\nu}$. Taking the infinite-volume scaling behavior $M(t, h) =h^{1/\delta}f(z)$ into account, one expects the combination 
\begin{equation}
M L^{\beta \delta/\nu} = Q_M(z, h L^{\beta \delta /\nu})
\label{eq:LOFSS}
\end{equation}
to be a universal function for a given value of $z$. This is indeed borne out by our results. In Fig.~\ref{fig:orderFVscaling}, the deviation from the infinite-volume behavior for the order parameter as a function of $h$ is readily apparent (left).
For large values of $h$, where the correlation length is small, the results for different volume sizes all approach
the infinite-volume limit. After the finite-size rescaling, the curves for the larger volume sizes once again are on top of each other (right), which clearly demonstrates the scaling behavior.
For the susceptibility $\chi$, the finite-size scaling behavior is shown in Fig.~\ref{fig:suscFVscaling}. For volume sizes  $10$ fm $ \le L \le 100$ fm, the unscaled susceptibilities differ by two orders of magnitude, but agree within $10 \%$ after rescaling.

For small volume sizes (for our choice of parameters $L \le 10$ fm), deviations from the leading-order scaling behavior become apparent. These deviations can be explained from an RG analysis of finite-size scaling beyond leading order, which takes irrelevant operators into account. Eq.~(\ref{eq:LOFSS}) then reads more fully with these corrections
\begin{equation}
M L^{\beta/\nu} = Q^{(0)}_M(z, hL^{\beta \delta/\nu}) + \frac{1}{L^\omega} Q^{(1)}_M(z, hL^{\beta \delta/\nu}) + \ldots,
\end{equation}
where the exponent $\omega$ is associated with the first irrelevant operator at the critical point. From a global fit to our results, we simultaneously determine the coefficient functions and the exponent $\omega$. Our result $\omega = 0.74(4)$ is in good agreement with $\omega = 0.7338$, which was obtained in \cite{Litim:2001hk} from an RG  fixed-point analysis carried out in infinite volume in the same RG scheme.
The results for the coefficient functions $Q^{(0)}_M(z, hL^{\beta \delta/\nu})$ and $Q^{(1)}_M(z, hL^{\beta \delta/\nu})$ determined from the volume range $10$ fm $ \le L \le 100$ fm are quite robust and almost unchanged when the analysis is extended to also take the region down to $L = 1$ fm into account, even though for our current choice of parameters the deviations from the leading-order scaling are already quite sizeable below $10$ fm.    

An important conclusion from this is that for small volumes, sub-leading corrections to the scaling behavior can become so large that the leading-order scaling analysis is no longer sufficient. While these results have presently been obtained only for a particular choice of parameters and the volume sizes cannot be immediately equated with those of QCD lattice calculations, they still suggest that scaling deviations play a role for small volumes in current lattice simulations.

\section{Conclusions}

In the work presented, we have investigated finite-size scaling in the O(N) universality class with non-perturbative Renormalization Group methods. For the O(4) model in $d=3$, the critical exponents and the scaling functions have been obtained, and the infinite-volume scaling behavior has been used to validate these results.

Adopting the RG scheme to calculations in finite volume, we have further demonstrated that finite-size scaling behavior can be obtained in this framework. We have calculated finite-size scaling functions for the order parameter and the susceptibility for a very wide range of values for the temperature and the symmetry-breaking field and for a wide volume range. 
The deviations from the leading-order scaling behavior that we observe can be adequately explained by the scaling corrections expected from an RG analysis. However, since these corrections are sizable for small volume sizes, they counsel caution for an analysis that takes only the leading-order scaling behavior into account for lattice results from small volumes. 

The results that we have obtained can be compared directly to results from QCD lattice simulations
and can be used to check compatibility with O(4) scaling behavior.
An extension to the O(2) universality class and such a direct comparison are in preparation.

\acknowledgments
BK would like to thank T. Mendes and C. Pica for very useful discussions.
This work was supported by the Excellence Cluster "Structure and Origin of the Universe" and the Natural Sciences and Engineering Research Council 
of Canada (NSERC). TRIUMF receives federal funding via a contribution agreement 
through the National Research Council of Canada.

\end{document}